\documentclass[runningheads]{llncs}
\usepackage{graphicx}
\usepackage{amsmath,amsfonts}
\usepackage{array}
\usepackage[caption=false,font=normalsize,labelfont=sf,textfont=sf]{subfig}
\usepackage{textcomp}
\usepackage{stfloats}
\usepackage{url}
\usepackage{verbatim}
\def\BibTeX{{\rm B\kern-.05em{\sc i\kern-.025em b}\kern-.08em
    T\kern-.1667em\lower.7ex\hbox{E}\kern-.125emX}}

\usepackage{algpseudocode}
\usepackage{algorithm}

\algnewcommand\algorithmicforeach{\textbf{for each}}
\algdef{S}[FOR]{ForEach}[1]{\algorithmicforeach\ #1\ \algorithmicdo}

\begin{document}
\title{Federated Learning-Based Interference Modeling\\ for Vehicular Dynamic Spectrum Access\thanks{The work has been realized within project no. 2018/29/B/ST7/01241 funded by the National Science Centre in Poland.}}
\titlerunning{FL-Based Interference Modeling for VDSA}
%
\author{Marcin Hoffmann\inst{1} \and
Pawel Kryszkiewicz\inst{1} \and
Adrian Kliks\inst{1}}

\authorrunning{M. Hoffmann et al.}
%
\institute{Institute of Radiocommunications, Poznań University of Technology, Polanka 3, 61-131 Poznań, Poland\\ \email{marcin.ro.hoffmann@doctorate.put.poznan.pl}}
\maketitle              
\begin{abstract}
A platoon-based driving is a technology allowing vehicles to follow each other at close distances to, e.g., save fuel. However, it requires reliable wireless communications to adjust their speeds. Recent studies have shown that the frequency band dedicated for vehicle-to-vehicle communications can be too busy for intra-platoon communications. Thus it is reasonable to use additional spectrum resources, of low occupancy, i.e., secondary spectrum channels. The challenge is to model the interference in those channels to enable proper channel selection. In this paper, we propose a two-layered Radio Environment Map (REM) that aims at providing platoons with accurate location-dependent interference models by using the Federated Learning approach. Each platoon is equipped with a Local REM that is updated on the basis of raw interference samples and previous interference model stored in the Global REM. The model in global REM is obtained by merging models reported by platoons. The nodes exchange only parameters of interference models, reducing the required control channel capacity. Moreover, in the proposed architecture platoon can utilize Local REM to predict channel occupancy, even when the connection to the Global REM is temporarily unavailable. The proposed system is validated via computer simulations considering non-trivial interference patterns.

\end{abstract}
\section{Introduction}
A platoon-based driving~\cite{Qin2018,MAITI20171,Bonnet2000} is a technology allowing vehicles to follow each other at short distances, i.e., of a few meters~\cite{tsugawa2010}. There are several significant benefits originating from this innovative approach. First, due to short inter-car distances, the road capacity can be improved by placing more vehicles on the road~\cite{vanArem2006}. Second, vehicles following each other benefit from better aerodynamic conditions. This further implies fuel savings, and reduction of emitted pollution~\cite{chan2012}. Most important, those gains from the platoon-based driving pattern are achieved with no need for additional road infrastructure, e.g., expensive and time-consuming building of additional road lanes. One, typically first, vehicle in the platoon is a so-called platoon leader. Depending on the concept, it can be either an autonomous vehicle or a vehicle with a human driver~\cite{fernandes2012,robinson2010operating}. In most cases, the platoon leader is responsible for the platoon behavior, and other vehicles not only adjust their speed to the platoon leader but also follow its maneuvers, e.g., line change. However, it is a non-trivial task, especially, when the platoon leader is forced to sudden breaking. Research has shown that for the purpose of safe driving while maintaining short inter-vehicle distances, it is not enough to rely on the onboard sensors or cameras~\cite{xu2014}. Additionally, reliable wireless communication must be established between the vehicles within the platoon to exchange control information, e.g., messages related to speed adjustment~\cite{thunberg2019}. There are several protocols already defined that are aimed at wireless communications between vehicles (known as a vehicle-to-vehicle scheme, V2V), between vehicles and network infrastructure (called vehicle-to-infrastructure communications, V2I), and between vehicles and pedestrians (named vehicle-to-pedestrians V2P)~\cite{abboud2016}. These are collectively referred to as vehicles-to-everything (named widely V2X). In general, the idea of V2X communications can be realized either in a centralized way with the use of cellular technology (i.e., Cellular-V2X, C-V2X~\cite{3gpp.37.985,harounabadi2021}), or in a distributed fashion by means of, e.g., Dedicated-Short-Range-Communication (DSRC)~\cite{802.11pSpec2010}. The DSRC solution can utilize 802.11p and Wireless Access in Vehicular Environment (WAVE) standards, in physical, and medium access layers, respectively~\cite{Abdelgader2014ThePL}. Although there are dedicated frequency bands for both DSRC and C-V2X, these might not be sufficient when road traffic is high. Some research has shown that with a growing number of vehicles utilizing DSRC interference in this part of the spectrum will drastically grow, decreasing the reliability of the wireless communications~\cite{REDDYG2018720,bohm2013}. Similar results were observed for C-V2X~\cite{wang2018}. The low reliability is caused by the fact that under the high level of interference the DSRC and C-V2X-dedicated frequency bands will not be able to ensure the capacity of the wireless channel required by the platoon. This would further imply a need to increase the inter-vehicle spaces within the platoon to prevent potential intra-platoon crashes causing e.g., lower fuel savings. This contradicts the overall concept of a platoon-based driving pattern discussed above.  

Therefore, it seems reasonable to look for alternative frequency bands where intra-platoon communication can be offloaded. Although almost all frequency bands are assigned to some wireless systems, the majority of these spectral resources are underutilized~\cite{kliks2013}. However, the frequency availability can vary with location or time. A platoon can potentially use a frequency band if it is not used at a given time in a given location. This requires detecting the radio activities of the so-called primary users (PUs), who are licensed to transmit in a particular frequency band. If PUs activity is not detected in this frequency band, this part of the spectrum can be opportunistically used by other systems. This approach is known in the literature as a Dynamic Spectrum Access (DSA)~\cite{zhao2007}, and when it relates to V2X communications, the Vehicle DSA (VDSA) scenario is then considered~\cite{chen2012vehicular}. In our work, we concentrate on the latter scheme.

Detection of the PU's signal presence may be done by means of the so-called spectrum sensing (such as energy detection). However, stand-alone spectrum sensing has limited performance. In addition, knowledge about the radio environment is out-dates fast as a result of vehicle movement in VDSA. An entity that can be used to retain and aggregate knowledge about the radio environment is a database called a Radio Environment Map (REM)~\cite{yilmaz2013}. The REM stores, updates, and process location-dependent information about the present PU's signal, which from the perspective of the unlicensed, so-called secondary user (SU) may be treated as interference. Such processing of historical data about interference possibly from many devices provides more insight into the availability of secondary spectrum channels. Moreover, please note that REMs may save also information about other ongoing SU transmissions.

The main challenge while utilizing a REM for VDSA is accurate modeling of interference caused by PU's signals along the route of the vehicle. In this context, the state-of-the-art models of interference utilized by REM are very simple, i.e., they characterize observed PU's signal power in a given location by its first-order statistics - the mean received power of unwanted signal ~\cite{kliks2017,yin2012,beek2012}. Such a simple approach was reliable enough when DSA was applied to the terrestrial television band, as the transmitted digital terrestrial television (DTT) signals are stable over frequency and time. However, in the general case, especially when unlicensed bands are considered, the PUs can utilize various algorithms for medium access controls and different transmission schemes. In such situations, more sophisticated models must be used to deal with complex interference distributions. This observation fits well with the considered VDSA scheme, where the observed unwanted signal distributions vary as a function of the road (location) and time. The creation of stable and precise models of unwanted signals may be time-consuming, and in practice, only a limited number of interference samples can be collected at one time. Thus, proper mechanisms for updating interference models in REM must be proposed. These mechanisms should be designed so as to enable the simultaneous contribution of several data sources, e.g., multiple platoons. 

In this paper, to deal with potentially complex distributions of interference generated by PUs, a Gaussian Mixture Model (GMM)~\cite{reynolds2009gaussian} will be used. GMM can model a Probability Density Function (PDF) of interference samples power, as a weighted sum of Gaussian distribution PDFs. Based on the model a wireless channel capacity can be calculated and the optimal frequency obtained using the algorithm proposed by us in \cite{hoffmann2021v2v}. However, that paper did not deal with a merger of knowledge from many sources. As the GMM model of interference varies in space, we proposed to store GMM-location pairs in the REM. Moreover, we propose a two-layer architecture of REM dedicated for VDSA, consisting of a Global, and a Local REM respectively. In our approach platoons can download the estimated GMM from Global REM to its Local REM, and update Local REM during ride based on local spectrum sensing. Later on, the locally updated GMM models from potentially many platoons are sent back to the Global REM to be merged into a single model therein. Such an architecture is enabled via Federated Learning (FL)~\cite{konecny2016FL,liu2022}. 
We confirm the effectiveness of this scheme via computer simulations and compare it against local sensing.  
The local interference modeling, without REM utilization, provides much worse insight into the distribution of interference for the platoon. Moreover, we show that the more platoons contribute to the update of Global REM the higher accuracy interference models can be obtained.

The paper is organized as follows: In Sec.~\ref{sec:system_model} a stand-alone system model is described, i.e., without REM. The proposed two-layered REM architecture together with formulas for an update of both local and global models is presented in Sec.~\ref{sec:rem_and_fl_modeling}. The simulation setup is described in Sec.~\ref{sec:simulation_scenario}. The results are discussed in Sec.~\ref{sec:results}. Finally, the conclusions are formulated in Sec.~\ref{sec:conclusions}.

\section{Stand-Alone System Model} \label{sec:system_model}
\begin{figure}[!t]
\centering
\includegraphics[width=3.6in]{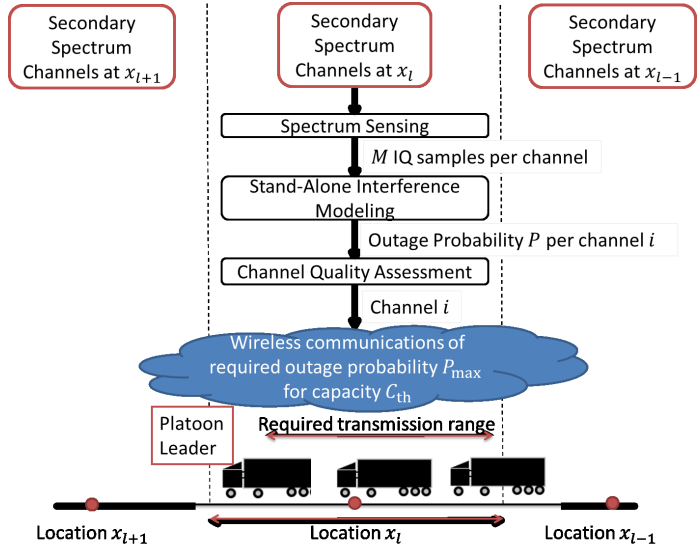}
\caption{Illustration of the stand-alone system model.}
\label{fig:system_model}
\end{figure}

Let us consider the generic case, where platoons offload a portion of their traffic from congested CV2X/DSRC bands to other frequency ranges, already occupied by the PU. All $U$ platoons apply the VDSA scheme trying to select the best frequency channels from all $\mathcal{I}$ candidate channels of bandwidth $B$. However, no additional edge intelligence, e.g., REM is used. 
A single platoon of index $u$ is formulated by $N_{u}$ vehicles, as depicted in Fig.~\ref{fig:system_model}. 
At a given time instance a platoon $u$ is located at position  $\mathbf{x}_l$ being a vector of geographical coordinates related to platoon location expressed in arbitrary chosen coordinates system e.g., earth-centered earth-fixed (ECEF). The behavior of the platoon, including maintaining proper inter-vehicle distances and management of intra-platoon communications, is realized by the first vehicle, which acts as a platoon leader. It sends proper management information to the remaining $N_{u}-1$ vehicles using wireless communications. To benefit from the platoon-based driving pattern, vehicles must drive at a very short distance. This implies wireless communication to meet high requirements on QoS. 
In our scenario, VDSA is assumed, so each platoon is looking for the secondary spectrum channel that would meet the platoon communication requirements. The selection of $i$-th frequency channel out of setting $\mathcal{I}$, consists of several steps (see Fig.~\ref{fig:system_model}), mainly spectrum sensing, interference modeling, and channel quality assessment These steps are described below  from the perspective of a single platoon, yet all $U$ platoons are performing them independently. 

\subsection{Capture of Samples}

The first step is to capture power samples from secondary spectrum channels available at location $\mathbf{x}_l$. We assume that every vehicle in the platoon can be potentially involved in this process. Thus samples can be simultaneously collected from several secondary spectrum channels. In each channel, a batch of $M$ in-phase and quadrature (IQ) received signal samples is collected in total.

\subsection{Channel Quality Assessment} \label{subsec:cqa}

Although in Fig.~\ref{fig:system_model} \emph{Interference Modeling} stands for the second step, it will be easier to understand the whole idea when the Channel Quality Assessment (CQA) would be described first. Moreover, some mechanisms used in interference modeling are related to the proposed CQA method.  

First, when discussing the CQA, the challenge is to choose the proper evaluation metric. From the perspective of overall communications reliability, one of the most important aspects is to monitor the latency~\cite{souza2020}. However, it is not straightforward to estimate it, as latency highly depends on the utilized radio access technology and medium access protocol. Both are in general not known when considering secondary spectrum channels, especially when many systems of different radio access technologies may operate. Moreover, according to the state-of-the-art models to assess latency, the knowledge of the arrival rate distribution of the incoming packets is necessary~\cite{zeng2019}. Its estimation is also case-dependent. 
On the other hand, a well-established metric is the Shannon capacity of a wireless channel being the upper bound estimate of the real system throughput. This metric is irrespective of the properties of the particular communication system, e.g., medium access control, and modulation schemes. Thus, in this paper, we follow this approach to assess channel quality. 

We compute the capacity of channel $i$ of bandwidth $B$ as a sum over $N_{\mathrm{f}}$ narrow sub-channels, which is motivated by the following observations. First, in the secondary spectrum, there may be potentially many interference sources of various frequency-dependent transmission characteristics, e.g., some of them may be narrow-band, and some wide-band. Second, in most of the contemporary communication protocols dedicated to V2V the multicarrier modulation scheme (Orthogonal-Frequency-Division-Multiplexing, OFDM) is applied, e.g., in IEEE 802.11p or CV2X using LTE or 5G networks. We consider the communication between the platoon leader and the last vehicle, as it is the most challenging because of the highest path loss. In consequence, at location $\mathbf{x}_l$ the Shannon capacity of the secondary spectrum channel $i$, between the two most distant vehicles within the platoon is given by:
\begin{equation} \label{eq:wifi_capacity}
c^{(i, \mathbf{x}_l)} = \frac{B}{N_{\mathrm{f}}} \sum_{f \in \mathcal{F}} \log_{2}\left(1 + \frac{P_{\mathrm{tx}}^{(i, \mathbf{x}_l)} \cdot H{(i, \mathbf{x}_l)}}{\sigma_{\mathrm{n}}^2 + I_{f}^{(i, \mathbf{x}_l)}}\right), 
\end{equation}
where $\mathcal{F}$ is a set of usable sub-channels (indexed by $f$), $H{(i,\mathbf{x}_l)}$ is a large scale channel gain (including transmit and receive antenna gains) at location $\mathbf{x}_l$ between the platoon leader and the last car. Next, $P_{\mathrm{tx}}^{(i, \mathbf{x}_l)}$ is the transmitter power per sub-channel, $\sigma_{\mathrm{n}}^2$ is the noise power over one sub-channel $f$, and $I_{f}^{(i, \mathbf{x}_l)}$ stands for the interference power on sub-carrier $f$. According to the field measurements, communication between vehicles following each other in the close distance, i.e., below $60$~m, is expected to be mainly under Line-Of-Sight (LOS) conditions for roof-top antennas~\cite{nilsson2017}. This observation can be applied to the communication between the platoon leader, and the last vehicle in a platoon consisting of a few vehicles, e.g. 3-4. As a result we expect channel to be relatively stable and flat. On the other hand interference in secondary spectrum potentially comes from the variety of sources having different transmission schemes. Therefore, our main focus is on the interference impact on the channel capacity.

While interference varies randomly, the capacity also becomes the random variable. From the perspective of the intra-platoon communication it is crucial to determine how likely is that channel capacity $c^{(i, \mathbf{x}_l)}$ falls below the acceptable level $C_{th}$, i.e., to derive the 
outage probability $\mathcal{P}(c^{(i,\mathbf{x}_l)}< C_{\mathrm{th}})$. Having \eqref{eq:wifi_capacity} as a starting point, in our previous work, we proposed to utilize Shannon capacity simplification proper for low Signal to Noise Ratio (SNR), in order to obtain the formula for the computation of the outage probability on the basis of interference distribution, and remaining transmission parameters, e.g., bandwidth, transmission power~\cite{hoffmann2021v2v}:
\begin{equation} \label{eq:outage_prob2}
    \mathcal{P}\left( \chi^{(i, \mathbf{x}_l)} < \ln{ \frac{\ln{2}\cdot C_{\mathrm{th}} \cdot N_{\mathrm{f}}}{\mathrm{B} \cdot P_{\mathrm{tx}}^{(i,\mathbf{x}_l)} \cdot H{(i,\mathbf{x}_l)}}}\right),
\end{equation}
where $\chi^{(i, \mathbf{x}_l)}$ is a random variable logarithm of aggregated interference and noise given by:
\begin{equation} \label{eq:agg_interference}
    \chi^{(i, \mathbf{x}_l)} = \ln \left(\sum_{f \in \mathcal{F}} \frac{1}{I_f^{(i, \mathbf{x}_l)}+\sigma^2_n} \right).
\end{equation}
With the use of the above equations quality of the available secondary spectrum, channels can be assessed, and the platoon leader can make the decision on the transmission channel to be in use. The detailed mathematical reasoning aimed at transformation of~\eqref{eq:wifi_capacity} into the~\eqref{eq:outage_prob2} can be found in~\cite{hoffmann2021v2v}. 

\subsection{Interference Modeling}

After the introduction of the CQA metric given by~\eqref{eq:outage_prob2}, it can be observed that this metric requires an accurate model of interference term~$\chi^{(i, \mathbf{x}_l)}$. The secondary spectrum channels can be occupied by many interference sources of different emission powers, diverse radio access technologies, various modulations, and not identical bandwidths. As a result the interference term $\chi^{(i, \mathbf{x}_l)}$ can follow a non-trivial, multi-modal distribution. It has been shown that such multi-modal distributions can be efficiently modeled with the use of the so-called Gaussian Mixture Model (GMM)~\cite{bishop2006}. The idea behind the GMM is to represent an arbitrary Probability Density Function (PDF), as a properly weighted sum of $J$ Gaussian distribution PDFs. In the considered case $\chi^{(i, \mathbf{x}_l)}$ is a one-dimensional random variable, thus GMM would be given by:
\begin{equation}\label{eq:gmm}
    p(\chi^{(i, \mathbf{x}_l)}) = \sum_{j=1}^{J}\pi_j \mathcal{N}(\chi^{(i, \mathbf{x}_l)} |\mu_j, \sigma_j),
\end{equation}
where $p(\chi^{(i, \mathbf{x}_l)})$ denotes the distribution of $\chi^{(i, \mathbf{x}_l)}$, $\pi_j$ is the $j$-th mixture component weight, i.e., the probability that $\chi^{(i, \mathbf{x}_l)}$ comes from the $j$-th mixture component. Next, $\mathcal{N}(\chi^{(i, \mathbf{x}_l)} |\mu_j, \sigma_j)$ is the conditional Gaussian distribution of $\chi^{(i, \mathbf{x}_l)}$, i.e., Gaussian distribution of $\chi^{(i, \mathbf{x}_l)}$, under assumption that it comes from the $j$-th mixture component with mean $\mu_j$ and standard deviation $\sigma_j$. To obtain parameters of a random distribution it is common to use a closed-form maximum likelihood estimator, i.e., closed form expressions are computed through maximization of the likelihood function. Unfortunately. in the case of GMM the log-likelihood function is a sum over the exponential functions, and no closed-form estimator exists~\cite{bishop2006}. Instead an iterative algorithm named Expectation Maximization (EM) is widely in use~\cite{dempster1977}. 

In this work the EM algorithm will be used to compute the parameters of GMM on the basis of $N_{\mathrm{s}}=\frac{M}{N_{\mathrm{f}}}$ samples of interference term $\chi^{(i, \mathbf{x}_l)}$. These samples are obtained from $M$ IQ samples captured each time any of the $U$ platoons visits location $\mathbf{x}_l$. The $M$ IQ samples are first divided into $N_{\mathrm{s}}$ non-overlapping segments, each of $N_{\mathrm{f}}$ samples. For each of these segments a Discrete Fourier Transform of size~$N_{\mathrm{f}}$ is applied. Next, the power samples are computed within each sub-channel. Finally, by applying~\eqref{eq:agg_interference} a single value of $\chi^{(i, \mathbf{x}_l)}$ is obtained for each of these segments. These~$N_{\mathrm{s}}$ samples of variable $\chi^{(i, \mathbf{x}_l)}$ constitutes an input to the EM algorithm. In the stand-alone approach discussed in this section, the platoon $u$ can rely only on the temporarily computed GMM to obtain the least occupied secondary spectrum channel $i$. In other words, the platoon senses the channel $i$ for the assumed period (obtaining $N_{\mathrm{s}}$ samples  of interference term $\chi^{(i, \mathbf{x}_l)}$), and generates the GMM interference model following the traditional EM approach. Once it is done, it decides on the spectrum occupancy and prospective capacity as described in Sec.~\ref{subsec:cqa}. 
We will refer to this schema as \emph{Stand-Alone Interference Modeling} (SAIM) throughout the rest of this paper, as it models the spectral usage characteristic without additional information provided by other entities, e.g., REM. 
We treat it as a reference solution. 

The SAIM approach has a significant drawback. The GMM model is computed only on the basis of a limited number of locally collected samples. To make the generated GMM reliable, the  simplest approach would be to set $N_{\mathrm{s}}$ to a large value. However, the primary goal of wireless communications within the platoon is to maintain safety on the road, especially when the inter-car distances are very short. Thus, it would be inefficient for the platoon to spend too much time on sensing, and thus $N_{\mathrm{s}}$ is expected to be low, creating a risk of inaccurate interference modeling through the GMM. In the following section (Sec.~\ref{sec:rem_and_fl_modeling}) we show how to overcome this issue by equipping the network infrastructure with REM modules utilizing a Federated Learning algorithm in order to enable continuous improvement of interference distribution models. 

\section{REM and Federated Learning for Interference modeling} \label{sec:rem_and_fl_modeling}

In order to overcome the drawbacks of the SAIM approach we propose to utilize historical knowledge gathered about the interference observed in the location $\mathbf{x}_l$ for better spectrum utilization estimation. We also propose the application of the federated learning approach for improving the interference modeling by exchanging interference awareness between all interested platoons.


\subsection{REM For Interference Modeling} \label{sec:rem_fl}

\begin{figure}[!t]
\centering
\includegraphics[width=2.8in]{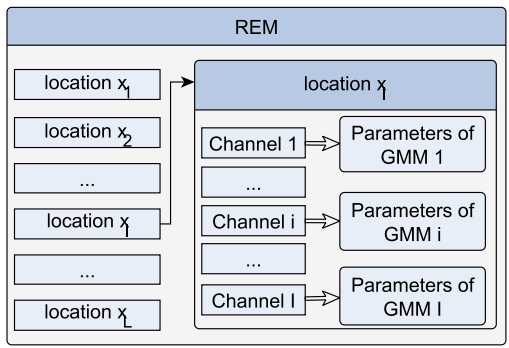}
\caption{The structure of data in REM.}
\label{fig:rem_data}
\end{figure}

\begin{figure}[!t]
\centering
\includegraphics[width=3.5in]{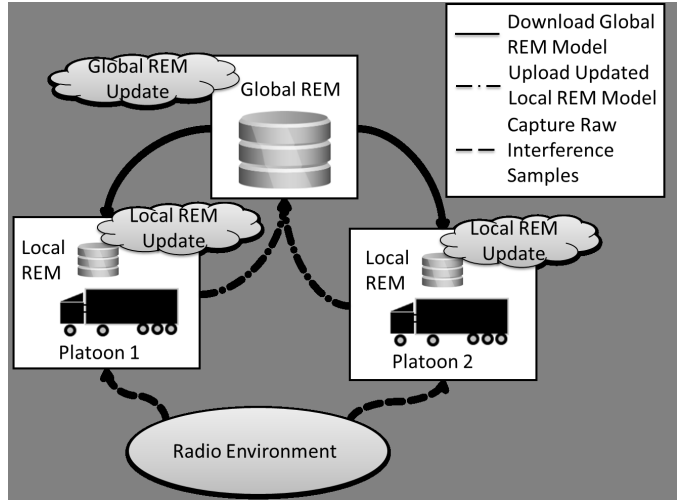}
\caption{Two-Layer REM architecture with FL cycle marked.}
\label{fig:layered_rem}
\end{figure}

\begin{figure}[!t]
\centering
\includegraphics[width=3.5in]{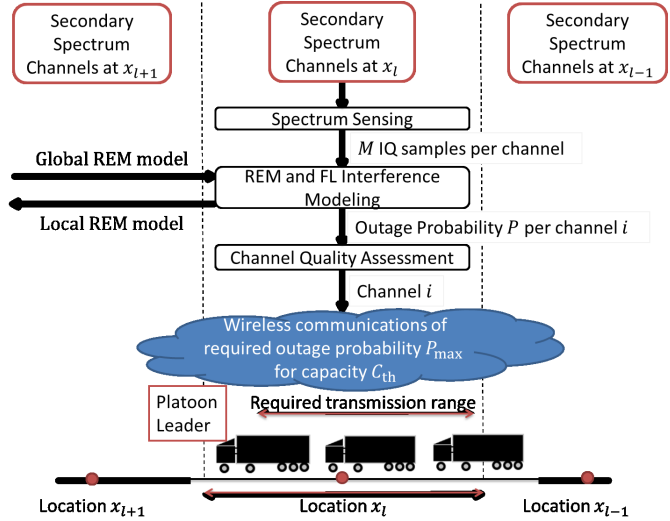}
\caption{Illustration of the REM and FL-based system model.}
\label{fig:system_model2}
\end{figure}

From the perspective of spectrum management and the best channel selection, the REM may be treated as an intelligent database containing location-dependent information about various signal sources (both, wanted and of interference type) observed in the wireless communication system~\cite{yilmaz2013}. The users can communicate with REM via a side link  to e.g. obtain information about the interference related to their positions or provide REM with their own measurement data. The entries in the database may have various forms, from raw measurements to some averaged figures of merit such as averaged received power in the band of interest. 
Although such REM databases may be applied to any frequency band and to any application, in the context of DSA the terrestrial television band was often in focus~\cite{SYB2018,katagiri2018,sroka2020}. In this frequency band, the interference is usually homogeneous, i.e., it has a stable average level over time and at a given location. In such a case, the interference can be easily modeled by the average power of the observed signal. However, as already mentioned, this idea can be  extended to build a REM database capable of modeling more complicated distributions, e.g., with interference power varying in time and frequency. Mainly, instead of average power, the parameters of GMMs related to the secondary spectrum channels $\mathcal{I}$ can be stored for each location $\mathbf{x}_l$. This structure of the data in REM is depicted in Fig.~\ref{fig:rem_data}. One can observe that each location is associated with the dictionary of tuples: channel and related GMM model parameters. 
From the architectural perspective, we propose a two-layered scheme, as depicted in Fig.~\ref{fig:layered_rem}, where Global REM and Local REMs are shown. 
Global REM is a part of the network infrastructure (belonging to, e.g., mobile network operator). It stores the global interference models, distributes them to the Local REMs, i.e., platoons, and is capable to combine them together in order to update its own global interference model. In turn, the Local REM is deployed and managed (updated) at each platoon $u$. Thus, in our case, in the first step, the Local REM will be populated by the data downloaded  from the Global REM. Next, these entries will be updated locally every time the platoon captures new $N_{\mathrm{s}}$ samples of interference term $\chi^{(i, \mathbf{x}_l)}$. Once processed locally, they will be used for updating the global REM, when the link to it is available. 
The idea behind the two-layered REM architecture aims at increasing the reliability of the system. While every platoon has its own Local REM, it can still use the interference models even when the side link to the Global REM is temporarily unavailable. Local REM is being updated on the basis of the raw samples, associated with the local environment, so it can reflect the current situation in the network and update the received model accordingly. As mentioned above, Local REM may also contribute to updating the Global REM models by sending back local observations or trained parameters. However, as the Local REM contains the whole GMM model it is enough to send only the model parameters to the Global REM instead of raw measurements. This results in significant traffic load reduction, i.e., order of tens parameters per GMM model vs hundreds of captured raw samples that have to be sent over the side link.
Such an approach is inspired by the popular federated learning scheme~\cite{yang2019,mcmahan2017,wang2019}, where the trained model details are exchanged between the central and surrounding nodes in order to better train the model and better reflect the local environment observed by each node. The procedure of secondary spectrum channel selection which utilizes interference modeling based on REM and FL is summarized in Fig.~\ref{fig:system_model2}. It can be seen that it is an extension of the state-of-the-art SAIM approach presented in Sec.~\ref{sec:system_model}. 

\subsection{Local REM Update} \label{subsec:local_update}
Let us denote the set of GMM parameters in Local REM related to the channel $i$, location $\mathbf{x}_l$, and platoon $u$ as $g_{u}(\mathbf{x}_l, i)$. These GMM parameters are obtained from the total number of samples equal to $N_{\mathrm{r},u}(\mathbf{x}_l, i)$. Every time when the location $\mathbf{x}_l$ is visited by the platoon $u$, only $N_{\mathrm{s}}$ new samples of interference term $\chi^{(i, \mathbf{x}_l)}$ per each secondary spectrum channel $i$ can be captured. These new samples are used to update the parameters of Local REM model $g_{u}(\mathbf{x}_l, i)$. The update procedure starts from computation of a temporal GMM model of $J$ components from those new $N_{\mathrm{s}}$ samples. The set of parameters of this temporal GMM model is denoted as $\tilde{g}_u(\mathbf{x}_l,i)$. 
The same number of GMM components $J$ for both $g_{u}(\mathbf{x}_l, i)$, and $\tilde{g}_u(\mathbf{x}_l,i)$. 
In general, $J$ can be established every time a new temporal GMM model of parameters set $\tilde{g}_u(\mathbf{x}_l, i)$ is built, by creating several models and comparing them against each other in terms of the chosen metric, e.g., Akaike Information Criterion (AIC), which is the function of model log-likelihood, and the number of parameters~\cite{akaike1974}. 
To reduce the number of GMM parameters to be sent between Local and Global REM we decided to use the shared-covariance variant of GMM, where a single variance is computed and shared among all $J$ components. 

 The parameters set of GMM model from Local REM $g_{u}(\mathbf{x}_l, i)$ consists of: shared variance $\sigma^2_u(\mathbf{x}_l,i)$, vector of means $\boldsymbol{\mu}_u(\mathbf{x}_l,i)=\{\mu_{u,j}(\mathbf{x}_l,i)\}_{j=1}^J$, vector of components proportions $\boldsymbol{\pi}_u(\mathbf{x}_l,i)=\{\pi_{u,j}(\mathbf{x}_l,i)\}_{j=1}^J$. The parameters set of temporal GMM model $\tilde{g}_u(\mathbf{x}_l,i)$ computed from the new $N_{\mathrm{s}}$ samples of interference term $\chi^{(i, \mathbf{x}_l)}$ is created by:  $\tilde{\sigma}^2_u(\mathbf{x}_l,i)$, $\tilde{\boldsymbol{\mu}}_u(\mathbf{x}_l,i)=\{\tilde{\mu}_{u,j}(\mathbf{x}_l,i)\}_{j=1}^J$, $\tilde{\boldsymbol{\pi}}_u(\mathbf{x}_l,i)=\{\tilde{\pi}_{u,j}(\mathbf{x}_l,i)\}_{j=1}^J$. For implementation purpose of combining together corresponding pairs of Local REM model parameters $g_{u}(\mathbf{x}_l, i)$ and temporal GMM model parameters $\tilde{g}_u(\mathbf{x}_l,i)$ we recommend to sort the vectors $\boldsymbol{\mu}_u(\mathbf{x}_l,i)$, $\boldsymbol{\pi}_u(\mathbf{x}_l,i)$, according to the $\boldsymbol{\mu}_u(\mathbf{x}_l,i)$ values, and vectors $\tilde{\boldsymbol{\mu}}_u(\mathbf{x}_l,i)$, $\tilde{\boldsymbol{\pi}}_u(\mathbf{x}_l,i)$ according to the $\tilde{\boldsymbol{\mu}}_u(\mathbf{x}_l,i)$. 

Now the parameters of the Local REM model $g_{u}(\mathbf{x}_l, i)$ are updated utilizing modified formula of incremental GMM learning~\cite{chen2012}:
\begin{equation} \label{eq:single_vehicle_update}
\begin{split} 
  \sigma_u(\mathbf{x}_l,i) &\leftarrow \frac{N_{\mathrm{th},u}(\mathbf{x}_l, i) \cdot \sigma_u(\mathbf{x}_l,i) + N_{\mathrm{s}} \cdot \tilde{\sigma}_u(\mathbf{x}_l,i)}{N_{\mathrm{th},u}(\mathbf{x}_l, i)+N_{\mathrm{s}}}, \\
    \mu_{u,j}(\mathbf{x}_l,i) &\leftarrow \frac{N_{\mathrm{th},u}(\mathbf{x}_l, i) \cdot \mu_{u,j}(\mathbf{x}_l,i) + N_{\mathrm{s}} \cdot \tilde{\mu}_{u,j}(\mathbf{x}_l,i)}{N_{\mathrm{th},u}(\mathbf{x}_l, i)+N_{\mathrm{s}}}, \\
    \pi_{u,j}(\mathbf{x}_l,i) &\leftarrow \frac{N_{\mathrm{th},u}(\mathbf{x}_l, i) \cdot \pi_{u,j}(\mathbf{x}_l,i) + N_{\mathrm{s}} \cdot \tilde{\pi}_{u,j}(\mathbf{x}_l,i)}{N_{\mathrm{th},u}(\mathbf{x}_l, i)+N_{\mathrm{s}}}, \\
    N_{\mathrm{r},u}(\mathbf{x}_l, i) &\leftarrow N_{\mathrm{r},u}(\mathbf{x}_l, i) + N_{\mathrm{s}},
\end{split}
\end{equation}
where $N_{\mathrm{th},u}(\mathbf{x}_l, i) = \min \{N_{\mathrm{r},u}(\mathbf{x}_l, i), k\cdot N_{s} \}$, and $k$ is a positive integer. The $N_{\mathrm{th}}$ parameter is defined so as to ensure that model can potentially follow the environment changes, it can be think of as a learning rate, i.e., how much impact on the model new data have. The larger the $k$ is the less impact the parameters of temporal GMM model $\tilde{g}_u(\mathbf{x}_l,i)$ have on the update of Local REM model parameters $g_{u}(\mathbf{x}_l, i)$. In the case when at a given platoon location $\mathbf{x}_l$ there is no information about interference model in REM, the values of GMM model parameters in Local REM are assumed to be initialized with set of temporal GMM model parameters $\tilde{g}_u(\mathbf{x}_l,i)$.    

\subsection{Global REM Update} \label{subsec:global_rem_update} 

In general, we can expect that multiple platoons can follow the same or partially overlapping route. A platoon $u$ will update its Local REM, following the procedure described in Sec.~\ref{subsec:local_update}. Suppose there are $U$ platoons that updated their Local REMS and send updated models back to the Global REM. The question arises of how to combine the $U$ models from Local REMs into the Global REM model. One could notice that this concept fits well into the idea of FL, where multiple clients train their local models, which are sent to server and processed to update the global model~\cite{abdulrahman2021}.  
Thus, in order to update the Global REM, We have decided to implement the state-of-the-art FL algorithm named FedAvg~\cite{mcmahan2017communicationefficient}.  The FedAvg formula for updating the Global REM is defined as a weighted average of the parameters from the Local REMs.
The detail update rule of GMM model parameters in Global REM related to channel $i$, and location $\mathbf{x}_l$ is presented below:  
\begin{equation}\label{eq:fed_avg_detail}
\begin{split}
\sigma_g(\mathbf{x}_l,i) &\leftarrow \frac{\sum_{u=1}^U N_{\mathrm{th},u}(\mathbf{x}_l, i)\cdot \sigma_u(\mathbf{x}_l,i)}{\sum_{u=1}^U N_{\mathrm{th},u}(\mathbf{x}_l, i)} \\
\mu_{g,j}(\mathbf{x}_l,i) &\leftarrow \frac{\sum_{u=1}^U N_{\mathrm{th},u}(\mathbf{x}_l, i)\cdot \mu_{u,j}(\mathbf{x}_l,i)}{\sum_{u=1}^U N_{\mathrm{th},u}(\mathbf{x}_l, i)} \\
\pi_{g,j}(\mathbf{x}_l,i) &\leftarrow \frac{\sum_{u=1}^U N_{\mathrm{th},u}(\mathbf{x}_l, i)\cdot \pi_{u,j}(\mathbf{x}_l,i)}{\sum_{u=1}^U N_{\mathrm{th},u}(\mathbf{x}_l, i)},
\end{split}
\end{equation}
where $\sigma_g(\mathbf{x}_l,i)$ is the  standard deviation shared among all GMM components, $\mu_{g,j}(\mathbf{x}_l,i)$ is the mean of $j$-th GMM component, and $\pi_{u,j}(\mathbf{x}_l,i)$ is the proportion of $j$-th GMM component.  

\subsection{Federated Learning Cycle}

The procedure of Local REM update described in Sec.~\ref{subsec:local_update}, together with the procedure of Global REM update described in Sec.~\ref{subsec:global_rem_update} would follow each other in a cyclic manner in order to learn Global REM with interference distributions. This constitutes an FL cycle, that according to the principles defined in~\cite{abdulrahman2021}, can be specified as shown in Algorithm~\ref{alg:fl}. This FL cycle can be visible also in the proposed two-layer REM architecture in Fig.~\ref{fig:layered_rem}. 

\begin{algorithm} \caption{FL Cycle for Update of Global REM}
\label{alg:fl}
\begin{algorithmic}[1]

\State Distribute parameters of Global REM model among $U$ platoons in the location $\mathbf{x}_l$ 

\ForEach{platoon $u \in U$}
    \ForEach{channel $i \in \mathcal{I}$}
    \State Capture $M$ new IQ interference samples
    \State Obtain $N_{\mathrm{s}}$ samples of interference term $\chi^{(i, \mathbf{x}_l)}$  
    \State Compute set of temporal GMM model 
    \Statex \hspace{1cm} parameters $\tilde{g}_u(\mathbf{x}_l,i)$ from $N_{\mathrm{s}}$ aggregated 
    \Statex \hspace{1cm} interference samples using EM algorithm
    \State {Update parameters of Local REM model 
    \Statex \hspace{1cm} $g_u(\mathbf{x}_l,i)$  with the use of eq.~\eqref{eq:single_vehicle_update}}
    \EndFor
\EndFor
\State Send sets of Local REM's models parameters $g_u(\mathbf{x}_l,i)$ from $U$ platoons to Global REM
\State Update Global REM following the eq.~\eqref{eq:fed_avg_detail}

\end{algorithmic}
\end{algorithm}

\section{Simulation Scenario} \label{sec:simulation_scenario}

\begin{figure}[!hbt]
\centering
\includegraphics[width=4.0in]{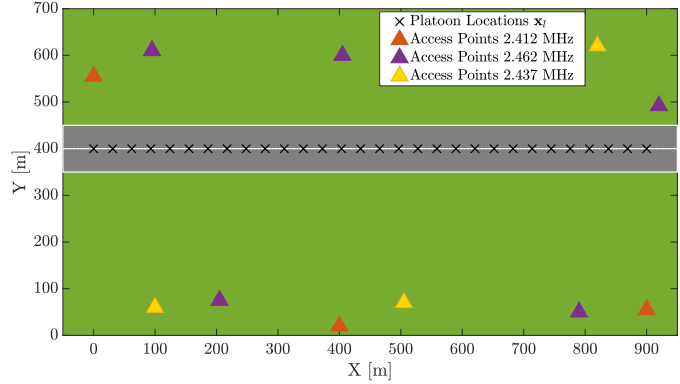}
\caption{Simulation scenario with platoon locations (black "x"), and access points locations (orange, purple, and yellow triangles respectively)}
\label{fig:scenario}
\end{figure}

Training of the GMM models requires a lot of data to be processed, especially when multiple platoons are considered. To guarantee the reliability of the model, a dedicated and detailed simulation environment has been built in order to generate interference samples along with the consecutive platoon locations. We consider three non-overlapping Wireless Local Area Network (WLAN) channels having $2.412$~GHz, $2.437$~GHz, and $2.462$~GHz center frequencies, respectively. At the same time, they formulate a set of available secondary spectrum channels $\mathcal{I}$, for which we model the interference using the proposed algorithms. Choice of this frequency band (i.e. ISM band) is motivated by the fact that there are many devices of potentially various medium access control algorithms, and modulations transmitting therein. As a result, non-trivial interference distributions are expected to be observed, which makes the modeling process challenging. 
The set of platoon locations $\mathbf{x}_l$ together with the deployment of interfering access points is depicted in Fig.~\ref{fig:scenario}. We are considering a 1 km long fragment of the route, which is split into 30 equally spaced segments (platoon locations) $\mathbf{x}_l$. Moreover, there are in total 11 wireless access points deployed along the considered platoon route. They generate interference over 3 different radio channels that correspond to orange, purple, and yellow triangles, respectively.

Every time a platoon $u$ visit location $\mathbf{x}_l$, a batch of $N_{\mathrm{s}}$ new aggregated interference samples is calculated according to~\eqref{eq:agg_interference}.  We assume bandwidth $B=10$~MHz, number of sub-channels $N_{\mathrm{f}}=64$, and number of used sub-channels $|\mathcal{F}|=48$ with spacing set to $156.3$~kHz, as defined in the 802.11p specification~\cite{Abdelgader2014ThePL}. 
To generate interference from the access point a four-state model of spectrum occupancy is used~\cite{geirhofer2006}. In this model idle state between acknowledgment message and transmission of new data, the block is modeled by an exponential distribution of traffic rate parameter $\lambda=0.0054$~ms$^{-1}$. The remaining busy states including the transmission of data and acknowledgment message last jointly 0.81~ms under the assumption of 512-byte packet size. To compute the large-scale fading coefficient of the radio channel between an access point and a platoon, a two-slope model proper for vehicular communications is used~\cite{Cheng_5_9GHz_Channel_model}. The small-scale variations of the radio channel are generated using Rayleigh distribution, and power-delay profile proper for the scenario where the platoon route is crossing an urban area~\cite{etsi2019channel}. Finally, the Gaussian-distributed thermal noise of power proper for the temperature of $20^\circ$~C is considered. 
 
The considered location can be simultaneously visited by the $U \geq 1$ platoons. In each platoon, the distance between the platoon-leader and the last $N_{\mathrm{u}}$-th vehicle is equal to $50$~m. It corresponds to a 3-truck platoon. It is the minimum number of vehicles that are required to capture interference samples from all sets of three secondary spectrum channels~$\mathcal{I}$. Transmitted power per subcarrier $\mathrm{P_{\mathrm{tx}}}$ value is the maximum allowed in the 2.4~GHz band~\cite{etsiWlan} and set to $3.19$~dBm, including antenna gains. Finally, 3~Mbit/s is set to be the desired capacity $C_{\mathrm{th}}$. It is the lowest supported bitrate in 802.11p, claimed to be used for emergency messages~\cite{campolo2013}.
 
The simulation parameters are summarized in Tab.~\ref{tab:scenario_parameters}.
 \begin{table}[!t]
\renewcommand{\arraystretch}{1.3}
\caption{Parameters of Simulation Scenario 
}
\label{tab:scenario_parameters}
\centering
\begin{tabular}{|c|c|}
\hline
Parameter & Value \\
\hline
number of vehicles in a platoon $N_\mathrm{u}$& 3 \\
platoon-leader to $N_{\mathrm{u}}$-th vehicle distance & 50~m\\
distance traveled by the platoon& 1~km \\
number of the platoon location & 30\\
number of access points & 11 \\
secondary spectrum channels $\mathcal{I}$ & $2.412$, $2.437$, $2.462$~GHz \\
secondary channel bandwidth $B$ & $10$~MHz \\
transmitted power per sub-carrier $\mathrm{P_{\mathrm{tx}}}$ & $3.19$~dBm \\ 
desired capacity $C_{\mathrm{th}}$ & $3$ Mbit/s \\
number of sub-channels $\mathrm{N_f}$ & 64 \\
number of usable sub-channels $|\mathcal{F}|$ & 48 \\
traffic rate parameter $\lambda$&  $0.0054$~ms$^{-1}$\\

\hline
\end{tabular}
\end{table}
\section{Results} \label{sec:results}

In this Section the evaluation of the algorithms proposed in Sec.~\ref{sec:system_model}, and Sec.~\ref{sec:rem_and_fl_modeling} is performed using the simulation environment described in Sec.~\ref{sec:simulation_scenario}.

\subsection{Baseline Model}

\begin{figure}[!t]
\centering
\includegraphics[width=3.9in]{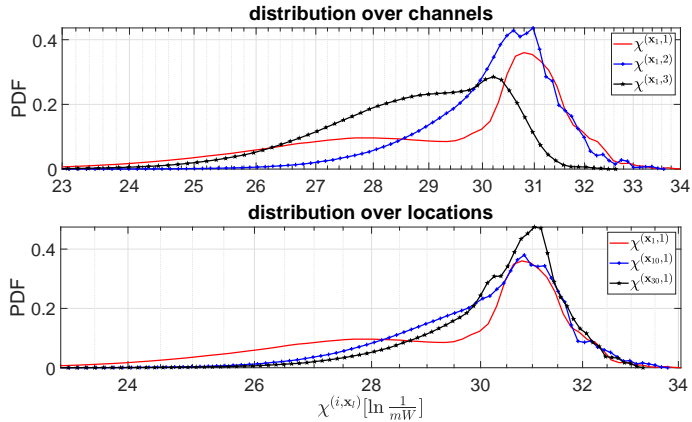}
\caption{Distribution of the aggregated interference $\chi^{(\mathbf{x}_l,i)}$ over different channels and platoon locations}
\label{fig:interf_distribution}
\end{figure}

\begin{figure}[!t]
\centering
\includegraphics[width=3.9in]{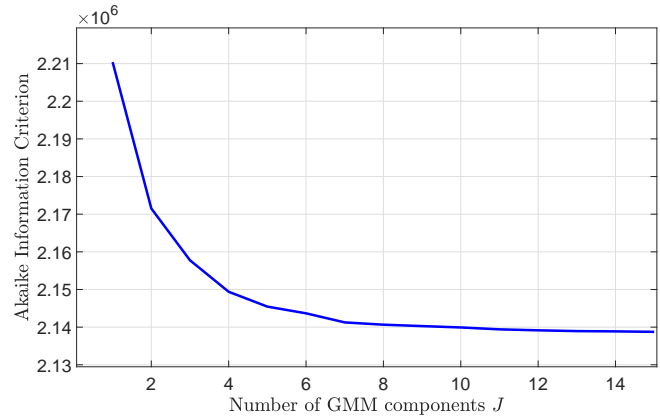}
\caption{Average AIC related to the GMM models of $J$ components. There are 90 training data sets considered. }
\label{fig:aic}
\end{figure}

At the first stage, we have analyzed an interference distribution $\chi^{(i,\mathbf{x}_l)}$ on the basis of large batch of captured samples, i.e., $M=39321600$ IQ samples, that corresponds to the $N_{\mathrm{s}}=614400$ samples of aggregated interference. This reflects a scenario where all data captured by the platoons were reported to REM, and once the statistically large number of samples has been collected, the proper GMM models can be computed. To prove our claim that interference can follow non-trivial, multi-modal distributions, there are representative examples of probability density functions related to $\chi^{(i,\mathbf{x}_l)}$ distributions  depicted in Fig.~\ref{fig:interf_distribution}. As it can bee observed, the distributions are complex, and they are varying over locations, and between channels.
On the basis of these results, GMM models have been created for every secondary spectrum channel in each of the 30 platoon locations. They will serve as baseline models for evaluation of the proposed algorithms: SAIM, and based on REM and FL. Set of parameters of these GMM models are denoted as $g_{\mathrm{B}}(\mathbf{x}_l,i)$. In order to determine the number of components $J$ for the GMMs, we have built baseline models of $J \leq 15$ components on the basis of $90$ data sets, i.e., 30 locations and 3 frequencies. The $g_{\mathrm{B}}(\mathbf{x}_l,i)$ models are evaluated in terms of their AIC.  The average AIC is depicted in Fig.~\ref{fig:aic}. It can be seen that after the number of GMM components is above $J=7$ the AIC remains stable, and the observed AIC improvement is very small. Thus, we decided to  fix this number, i.e., $J=7$ in the following simulations. Clearly, in practice, the value of $J$ has to be adjusted to each situation, based on observed and sensed interference sources, and based on the length of the considered road fragment. 

\subsection{Single Platoon Scenario}
\begin{figure}[!t]
\centering
\includegraphics[width=3.9in]{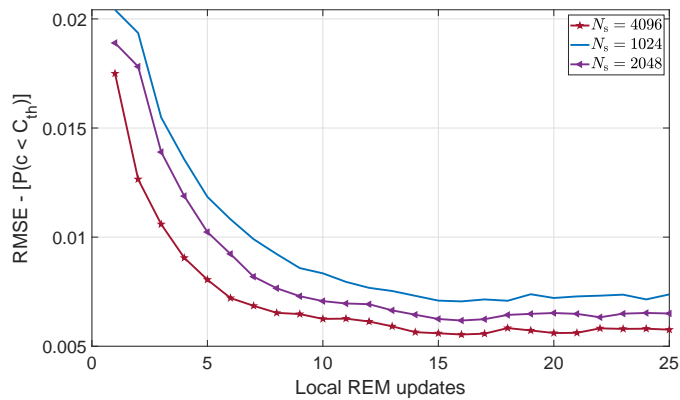}
\caption{ RMSE computed between the outage probability estimated with the use of updated Local REM and the baseline model, for $k=5$. Results are averaged over 50 simulations.}
\label{fig:sort_local_batch}
\end{figure}

We first evaluate the accuracy of the proposed method of interference modeling in the  scenario where the number of the platoons that simultaneously travel the considered route is set to one, i.e., $U=1$. In this scenario, our main aim is to evaluate the algorithm of the Local REM update as described in Sec.~\ref{subsec:local_update}. In one simulation run, the platoon travels the same route 25 times, i.e., Local REM is updated 25 times at each location $\mathbf{x}_l$. To smooth the results, up to 50 simulation runs are performed. There are two parameters to be evaluated: the number of captured aggregated interference samples $N_{\mathrm{s}}$,  and the learning rate $N_{\mathrm{th},u}(\mathbf{x}_l, i)$ driven by the parameter $k$, as defined by~\eqref{eq:single_vehicle_update}. First, $k$ is arbitrary set to 5, and the impact of the number of captured aggregated interference samples $N_{\mathrm{s}}$ on the Local REM update is under consideration. The Root-Mean-Square-Error (RMSE) computed between the outage probability estimated with the use of updated Local REM and the baseline model is depicted in Fig.~\ref{fig:sort_local_batch}.  
It can be seen that after about $15$ updates of Local REM, i.e., after the platoon traveled the considered route 15 times, all results tend to stabilize, and no further improvement of the RMSE could be observed. However, the level of stabilization depends on the number of captured aggregated interference samples $N_{\mathrm{s}}$. For the lowest $N_{\mathrm{s}}=1024$ RMSE is approximately equal to $0.007$, while for the $N_{\mathrm{s}}=4096$ is about $0.0055$, that is over 20\% improvement with $N_{\mathrm{s}}$ increase. Obviously, it can be further improved by increasing $N_{\mathrm{s}}$, but the cost to be paid is less time for the intra-platoon communications. On the other hand, the initial analysis not presented here shows that decreasing the $N_{\mathrm{s}}$ leads to computational instability while estimating the set of GMM model parameters~$\tilde{g}_u(\mathbf{x}_l, i)$.

\begin{figure}[!t]
\centering
\includegraphics[width=3.9in]{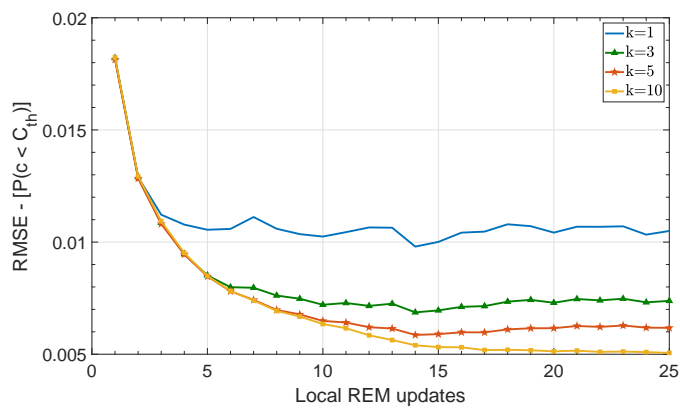}
\caption{ RMSE computed between the outage probability estimated with the use of Local REM and the baseline model, for $k=5$. Results are averaged over 50 simulations.}
\label{fig:sort_local_memory}
\end{figure}

Later on, we have investigated the impact of the parameter $k$ defining the learning rate on the Local REM updates. This time the number of collected aggregated interference samples $N_{\mathrm{s}}$ is set to the $4096$. The related RMSE between the outage probability computed with the use of Local REM and baseline model, respectively, is depicted in Fig.~\ref{fig:sort_local_memory}.
 
It can be seen that the higher the parameter $k$, the better the RMSE. However, one should note that in the considered case interference being produced by each of the access points is a  stationary random process. In the practical systems, it might not be true, thus it is reasonable not to choose very big $k$ allowing adaptation to the changing radio conditions, e.g., turn-off of an access point. Moreover, it can be seen that when $k=1$, results are unstable. It is because in this case newly computed GMM model possibly of low accuracy because of a limited number of samples $N_{\mathrm{s}}$, has the same weight as the GMM model stored in Local REM. After the stabilization of the results, RMSE values obtained for $k=10$ are approximately 5 times better than those for $k=1$. The final conclusion is that $k$ should be large enough to provide stable results, but small enough to follow the radio environment changes. 

\subsection{Multiple Platoons Scenario}

\begin{figure}[!t]
\centering
\includegraphics[width=3.9in]{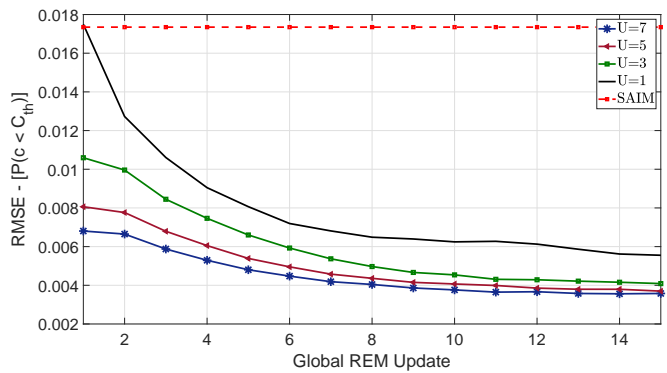}
\caption{ RMSE computed between the outage probability estimated with the use of Global REM and the baseline model, for $N_{\mathrm{s}}=4096$, $k=5$, and the varying number of platoons $U$. Additionally, the results of the SAIM scheme are included. Results are averaged over 50 simulations.}
\label{fig:no_rem}
\end{figure}

Having the algorithm of Local REM updates evaluated now is the time to move to the general case where $U\geq1$ platoons can capture interference samples simultaneously at a given location $\mathbf{x}_l$, and participate in the process of improving the Global REM model, following the proposed FL-based approach. In this scenario, $U$ platoons travel through the set of locations $15$ times. This procedure is repeated 50 times in order to average the results. Now, we set number of aggregated interference samples $N_{\mathrm{s}}$ to $4096$, and the $k$ parameter to $5$. Our focus is put on the impact of the number of platoons $U$ on the Global REM improvement. Similar to the previous plots, the outage probability computed with the Global REM and baseline models are compared in terms of RMSE in Fig.~\ref{fig:no_rem}. In addition, the plot contains a result that refers to the scenario where the interference model is computed independently at each platoon only on the basis of the currently captured $N_{\mathrm{s}}$ samples, i.e., SAIM. The first observation is that benefits from the utilization of historical data in REM and FL are significantly compared to the SAIM scenario. In the case of a single platoon, $U=1$ RMSE after many REM updates is about 3 times lower than RMSE obtained for SAIM which is not improving as a result of lack of memory in REM. Moreover, in the case of simultaneous Global REM updates by $U=7$ platoons, a 4.5 fold reduction of the RMSE can be observed after 25 REM updates referring to the SAIM. Comparing the systems utilizing Global REM Updates, the biggest difference is between the single platoon scenario and the $U=3$ platoon scenario, i.e., about 27\% after 25 REM updates. While increasing further the number of simultaneously sensing platoons only a very little improvement can be observed, e.g., comparing between $U=5$ and $U=7$ the improvement is at the level of about 3\%.

\subsection{Channel Selection}
\begin{figure}[!t]
\centering
\includegraphics[width=3.9in]{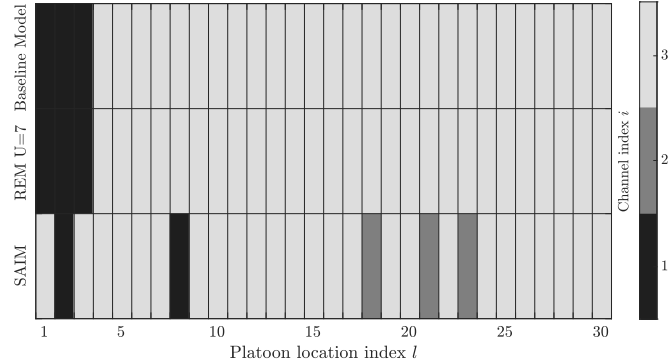}
\caption{ Index $i$ of channel offering the lowest outage probability in the consecutive platoon locations $\mathbf{x}_l$ computed on the basis of baseline, Global REM, and \emph{Spectrum Sensing} interference model.}
\label{fig:channel_selection}
\end{figure}

It is important to take into account that the proposed algorithm of interference modeling realized with the use of a REM and FL is not the target itself. It is a tool to enable the selection of a proper secondary spectrum channel, as described in Sec.~\ref{sec:system_model}. On the basis of the interference models stored in REMs, platoons can decide which channels should be chosen along their route, e.g., on the basis of Dijkstra's algorithm, as discussed in~\cite{hoffmann2021v2v}. In this paper, the simplest approach is evaluated, where a platoon chooses a wireless channel characterized by the lowest outage probability. In Fig.~\ref{fig:channel_selection} there is an index of channel offering the lowest outage probability in the consecutive platoon locations computed on the basis of the baseline model, the model utilizing REM (with Local and Global REM updates), and SAIM. We assume that the Global REM that is used is built from data captured by the $U=7$ platoons during 15 route travels, and the considered parameters are: $N_{\mathrm{s}}=4096$, and $k=5$. It can be seen that the channels marked as the best on the basis of the interference model from REM cover up with the ones computed using the baseline model in all cases, which is highly expected and proves the correctness of the proposed method. On the other hand, results based on SAIM correctly indicate the best radio channels only in 24 out of 30 platoon locations. It means that on the basis of only currently available interference samples the best secondary spectrum channel can be selected only in 80 \% of cases. It is important to note that such a mistake can have a significant impact on the quality of wireless communications between vehicles formulating a platoon. This can further imply decreased fuel saves or platoon crash in extreme case.

\section{Conclusion} \label{sec:conclusions}

In this paper, we proposed a two-layer architecture of REM, that together with FL and local spectrum sensing can be effectively used to provide platoons with an accurate model of interference. We have shown that the GMM accurately models the non-trivial distributions of interference in VDSA scenarios.
Through the extensive computer simulations studies, we came to conclusion that historical data from REM can much improve the accuracy of interference models used by the platoon to predict radio channel occupancy in relation to the state-of-the-art SAIM approach. Moreover, it has been observed, that due to utilization of the FL multiple platoons can contribute to the improvement of Global REM interference models even more. Up to $4.5$~fold reduction in channel capacity estimate RMSE was observed while comparing algorithm based on REM and FL against standard SAIM approach. It is important to notice that the proposed two-layer REM architecture have low requirements on the capacity of control channel. This is because only the GMM parameters are exchanged between the Local and the Global REM, instead of direct spectrum sensing results. Finally, equipping every platoon with Local REM ensures that channel capacity can be predicted even, when the side link to the Global REM is temporarily unavailable. What increases the system's reliability. In the future the measurement campaign can be conducted in order to verify the proposed algorithms in real-world scenario.
%
%
%
 \bibliographystyle{splncs04}
 \bibliography{bibliografia}

\end{document}